\documentclass[pra,twocolumn,preprintnumbers,showpacs,
nofootinbib,floatfix]{revtex4-1}

\usepackage{graphicx,bm,amsmath}
\usepackage{bm}
\usepackage{tikz}
\usepackage[sort&compress]{natbib}

\makeatletter

\def\graphicscale{\twocolumn@sw{0.30}{0.33}}
\def\graphicthreescale{\twocolumn@sw{0.30}{0.33}}

\makeatother

\begin{document}

\include{macros}

\title{NSPT estimate of the improvement coefficient $c_A$ to two loops}

 \author{Christian Torrero}

 \date{\today}

\begin{abstract} 

By using Numerical Stochastic Perturbation Theory (NSPT), we carry out a quenched two-loop computation
of the improvement coefficient $c_A$ associated to the isovector axial current. Within the Schr\"odinger Functional formalism,
we compute the bare PCAC quark mass $m$ and fix $c_A$ by requiring discretization corrections on $m$ to be of order $\emph{O}(a^2)$
in the lattice spacing $a$. 

\end{abstract}


\maketitle


\emph{Introduction -} After performing a set of numerical simulations of a generic field theory,
corresponding results become meaningful only if extrapolated to the physical point ($\Phi$). This means that, given simulation 
parameters like --- among others --- lattice spacing $a$, volume $V$ and, in case, quark masses $m_q$, the limits 
$a\rightarrow 0$, $V\rightarrow +\infty$ and $m_q \rightarrow m_q^{(\Phi)}$ have to be accurately and reliably computed. 
While these limits were thought to be hardly accessible for lattice QCD some time ago~\cite{Ukawa:2002pc}, a few decades of algorithmic 
developments have paved the way to controlling all the above-mentioned sources of systematic uncertainty. Nowadays, 
not only $N_f = 1+1+1+1$ QCD simulations with small lattice spacings, large boxes and quark masses close to the physical point
are being performed, but also challenging QED effects have started to be taken into account~\cite{Borsanyi:2014jba}.

At present, the standard setup for QCD simulations features the Hybrid Monte Carlo (HMC) algorithm~\cite{Duane:1987de}, usually supplemented
with other techniques --- like preconditioning~\cite{DeGrand:1990dk}, the Hasenbusch trick~\cite{Hasenbusch:2001ne}, multiple time-scale integration~\cite{Sexton:1992nu} and 
smearing~\cite{Albanese:1987ds,Hasenfratz:2001hp,Morningstar:2003gk,Capitani:2006ni}. In the framework of the HMC algorithm, 
such techniques are useful in that they allow for a larger value of the time-step in integrating the equations of motion (thus 
decreasing the autocorrelation among subsequent configurations), they reduce the condition number $\kappa(D)$ of the Dirac operator $D$ 
(speeding up the inversion of $D$ needed both within the HMC algorithm and at measure time) and they average out ultraviolet fluctuations, 
improving the signal-to-noise ratio (SNR).    

In this scenario, a technique often employed to reduce discretization effects is the so-called improvement, usually in a form following
the Symanzik programme~\cite{Symanzik:1983dc}. To understand how the latter works, it is useful to recall that the mean value $\overline{O}$ of a generic
continuum observable $O$ computed on the lattice reads
\begin{equation}
\label{Syma}
\overline{O}(a,V,m_q,\ldots) = \int DU D\psi D\overline{\psi}\ \!O_{Lat}\ \!e^{-S_{Lat}}\ ,
\end{equation}
where $O_{Lat}$ is the lattice counterpart of $O$, $S_{Lat}$ is the lattice QCD action (given by the sum of a gauge part $S_G$ and a fermionic part $S_F$) and where
the dependence of $\overline{O}$ with respect to simulation parameters has been made explicit.

\indent According to Symanzik, close to the continuum limit, $O_{Lat}$ can be Taylor-expanded in the lattice spacing as 
\begin{equation}
\label{Sym}
O_{Lat} = O + aO_1 + a^2O_2 + \ldots \ ,
\end{equation}
where $O_1$, $O_2$, $\ldots$ have to be interpreted as contributions stemming from operator insertions in the continuum and must have symmetry properties 
consistent with $O$. A similar expression holds also for the gauge and the fermionic action $S_G$ and $S_F$ entering $S_{Lat}$.

\indent Plugging said Taylor expansions with respect to $a$ into Eq.(\ref{Syma}) results in a similar expansion for $\overline{O}$ as well.
In the Symanzik improvement programme, the leading correction to $\overline{O}$ --- usually linear in $a$ as in Eq.(\ref{Sym}) --- can be cancelled by adding 
irrelevant terms to $S_G$, $S_F$ and to $O_{Lat}$\footnote{In the rest of the paper, an action/observable whose leading correction in $a$ is of order $a^j$ will 
be defined as $\emph{O}(a^{j-1})-$improved.}. In this way, the dependence of $\overline{O}$ with respect to $a$ is flattened and, consequently, larger values of the 
lattice spacing can be used to recover the continuum limit, thereby further reducing $\emph{k}(D)$ and increasing the SNR at the same time. 

In general, each irrelevant term is multiplied by its own coefficient that has to be appropriately tuned: its value can be determined either 
non-perturbatively or within perturbation theory (PT). 
Obviously, among these so-called improvement coefficients, the most important ones are those improving on $S_G$ and $S_F$ since they enter in the improvement 
procedure of any observable. When it comes to perturbative computations, the expansions of these coefficients in the bare coupling $g_0$ are usually known 
up to very low orders only, usually at one loop.

In this paper, we investigate whether Numerical Stochastic Perturbation Theory (NSPT)~\cite{DiRenzo:1994sy,DiRenzo:2004hhl} can be applied to compute 
the improvement coefficients in PT to orders higher than $g_0^2$. We want to clearly state that the present study essentially aims at being a 
\emph{proof of concept}, i.e., at assessing the feasibility --- or not --- of a similar NSPT computation in principle. Given such exploratory character, we 
tackle the simplest case possible, namely the two-loop computation of the improvement coefficient $c_A$ associated to the isovector axial current in the quenched
approximation. In spite of its seeming minor importance, a perturbative computation of $c_A$ to a given order $j$ allows for a determination to the same order
of the much more important coefficient $c_{SW}$, i.e., the improvement coefficient multiplying the irrelevant term improving on the fermionic action 
$S_F$~\cite{Sheikholeslami:1985ij} which will be introduced later. 

\emph{Lattice setup -} In this and the next section we outline our approach which is based on the Schr\"odinger Functional formalism~\cite{Sint:1993un,Luscher:1992an} 
and closely follows the strategy described in~\cite{Luscher:1996sc} and used in~\cite{Luscher:1996vw} to compute $c_{SW}$ and $c_A$ to one loop.
 
We simulate a four-dimensional lattice made up of $N_T\times N_S^3$ sites, each one labelled with integer coordinates $n=(n_0,n_1,n_2,n_3)$ 
varying in the intervals $[0,N_T-1]$ and $[0,N_S-1]$ along the time and spatial directions respectively. The lattice volume $V$ will therefore be equal
to $V=L_T\times L_S^3$, with $L_T = a(N_T-1)$ and $L_S = aN_S$. 

A generic gauge variable $U_{\mu}(n)$ --- with $\mu\in\{0,1,2,3\}$ --- belongs to the $SU(3)$ group and is associated to the link connecting site $n$ to site $n+\hat{\mu}$, 
$\hat{\mu}$ being a unit vector along direction $\mu$. Lattice group variables are related to their continuum counterparts $A_{\mu}(n)$ in the Lie algebra of $SU(3)$ through 
the equation $U_{\mu}(n)=exp[iA_{\mu}(n)]$. We stick to the usual convention according to which $U^{-1}_{\mu}(n) = U^{\dagger}_{\mu}(n)$.

Quark and antiquark degrees of freedom are Grassmann variables --- denoted as $\psi(n)$ and $\bar{\psi}(n)$ respectively --- 
associated to the lattice sites\footnote{Spin, colour and flavour indices are always left implicit for all fields, except where strictly needed.}. For simplicity, 
we assume the presence of $N_f$ mass-degenerate flavours though, in what follows, the fermionic action $S_F$ and its irrelevant term will always be written by taking 
into account only one flavour to ease the notation: it is understood that there are actually $N_f$ replicae of such operators. Anyway, it is worth stressing that, while the 
quenched approximation implies --- obviously --- that fermionic degrees of freedom play no role in updating the lattice configuration and that, consequently, the results of 
this paper have to be considered valid for $N_f=0$, the setup described in the next sections is such that $N_f$ never enters into play at measurement time as well, as long 
as the mass degeneracy holds.

While boundary conditions are periodic along the three spatial directions, they are of Dirichlet type in the time direction. In other words, by labelling 
a generic spatial direction with $k$ from now on, for the gauge fields the following equalities hold
\vspace*{0.25cm}
\begin{equation}
\left.U_k(n)\right|_{n_0=0} = W_k(\vec{n})\ , \ \ \ \left.U_k(n)\right|_{n_0=N_T-1} = W_k'(\vec{n})\ ,
\vspace*{0.15cm}
\end{equation}
\noindent with $\vec{n}=(n_1,n_2,n_3)$ and where $W_k(\vec{n})$ can be expressed in terms of a smooth, fixed field $C_k(\vec{n})$ as
\begin{equation}
\vspace*{0.25cm}
W_k(\vec{n}) = \mathcal{P}\ \!exp\bigg[a\!\!\int_0^1\!\!dt\ \!C_k(\vec{n}+a\hat{k}-ta\hat{k})\bigg]\ ,
\vspace*{-0.05cm}
\end{equation}
being $\mathcal{P}$ the path-ordering symbol. $\!W_k'(\vec{n})$ is parametrized by another field $C_k'(\vec{n})$ in an analogous way. 

With this setup, the lattice gauge action $S_G$ is given by the modified Wilson action
\begin{equation}
\label{SG}
S_G = \frac{1}{g_0^2}\sum_p \omega(p)tr\{1-U(p)\}\ ,
\end{equation}
where $U(p)$ is the product of the link variables around a lattice plaquette, the sum runs on all oriented plaquettes and the weights $\omega(p)$ are equal 
to $1$ for each plaquette, except for the spatial ones at $n_0=0$ and $n_0=N_T-1$ where $\omega(p)=\frac{1}{2}$. Due to the SF boundary conditions, the gauge action
$S_G$ in Eq.(\ref{SG}) with said values of the weights $\omega(p)$ is $\emph{O}(a)-$improved only at tree-level in PT. A version of $S_G$ that is $\emph{O}(a)-$improved at 
any perturbative order can be obtained by adding some boundary counterterms featuring their own improvement coefficients that have to be appropriately tuned. This whole 
procedure would eventually amount to a redefinition of the weights $\omega(p)$ close to the boundaries. In the present work, such $S_G-$related counterterms will be ignored 
because the observable that will be introduced and studied later on --- i.e., the bare PCAC quark mass --- is entirely fixed by a Ward identity and this peculiar property 
allows to neglect said counterterms.  

As for the fermionic degrees of freedom, after introducing the projectors $P_{\pm} = \frac{1}{2}(1\pm\gamma_0)$ --- $\gamma_0$ being a Euclidean Dirac matrix --- and some fixed 
Grassmann fields $\rho,\ldots,\bar{\rho}'$, their Dirichlet boundary conditions are given by
\vspace{0.35cm}
\begin{equation}
\label{DirF1}
\left.P_+\psi(n)\right|_{n_0=0}=\rho(\vec{n})\ , \ \ \ \ \left.P_-\psi(n)\right|_{n_0=N_T-1}=\rho'(\vec{n})\ ,
\vspace{0.25cm}
\end{equation}
\noindent for the quark fields and by
\vspace{0.35cm}
\begin{equation}
\label{DirF2}
\left.\bar{\psi}(n)P_-\right|_{n_0=0}=\bar{\rho}(\vec{n})\ , \ \ \ \left.\bar{\psi}(n)P_+\right|_{n_0=N_T-1}=\bar{\rho}'(\vec{n})\ ,
\vspace{0.25cm}
\end{equation}
\noindent for the antiquark fields. For consistency, quantities $P_-\rho,\ldots,\bar{\rho}'P_-$ must vanish.

\indent The unimproved fermionic action $S_F$ is given by
\vspace*{0.2cm}
\begin{equation}
\label{SF}
S_F = a^4\sum_{n}\bar{\psi}(n)(D+m_0)\psi(n)\ , 
\vspace*{0.1cm}
\end{equation}
$m_0$ being the bare quark mass and the Wilson-Dirac operator $D$ reading
\vspace*{0.2cm}
\begin{equation}
\label{SFunimp}
D = \frac{1}{2}\bigg[\gamma_{\mu}(\nabla_{\mu}^*+\nabla_{\mu})-a\nabla_{\mu}^*\nabla_{\mu}\bigg]\ ,
\vspace*{0.1cm}
\end{equation}
where repeated indices are summed and $\gamma_{\mu}$'s are Euclidean Dirac matrices.
Covariant derivatives in Eq.(\ref{SFunimp}) are defined as
\vspace*{0.2cm}
\begin{eqnarray}
\label{covdev}
\!\!\!\nabla_{\mu}\psi(n) &=&\frac{1}{a}\big[\lambda_{\mu}U_{\mu}(n)\psi(n+\hat{\mu}) - \psi(n)\big] \ , \nonumber\\
\!\!\!\nabla_{\mu}^*\psi(n) &=&\frac{1}{a}\big[\psi(n) - \lambda_{\mu}^{-1}U_{\mu}^{-1}(n-\hat{\mu})\psi(n-\hat{\mu})] \ ,\\  
&&\nonumber
\vspace*{-0.0cm}
\end{eqnarray}
$\lambda_{0}=1$ and $\lambda_{k}=exp(ia\theta_{k}/L_S)$ being phase factors (with $-\pi<\theta_{k}\le\pi$). For simplicity, all three angles $\theta_k$
will be set to the same unique value $\theta\ne0$. 

Strictly speaking, Eq.(\ref{SF}) holds in an infinite volume. However, it remains valid also in the present setup --- i.e., a box of finite size with Dirichlet boundary
conditions --- provided some technical conventions are assumed: the interested reader can find more details in subsection $4.2$ of~\cite{Luscher:1996sc}. We tacitly take 
such conventions for granted and carry on with Eq.(\ref{SF}) in combination with said lattice topology.


\indent The leading discretization correction to $S_F$ is linear in $a$ and, as mentioned in the introduction, an $\emph{O}(a)-$improved fermionic 
action $S_F^{imp}$ can be obtained by adding to Eq.(\ref{SF}) an irrelevant term $\delta S_V$, i.e.,
\vspace{0.2cm}
\begin{equation}
S_F^{imp} = S_F + \delta S_V\ .
\vspace{0.2cm}
\end{equation}
$\delta S_V$ is usually referred to as clover term and its expression is given by~\cite{Sheikholeslami:1985ij} 
\vspace{0.2cm}
\begin{equation}
\label{clover}
\delta S_V = a^5c_{SW}\sum_{n_0=1}^{N_T-2}\sum_{n_1,n_2,n_3=0}^{N_S-1}\bar{\psi}(n)\frac{i}{4}\sigma_{\mu\nu}F_{\mu\nu}(n)\psi(n)\ ,
\vspace{0.2cm}
\end{equation}
where $\sigma_{\mu\nu} = \frac{i}{2}[\gamma_{\mu},\gamma_{\nu}]$ and
\vspace{0.2cm}
\begin{equation}
\label{Fmunu}
F_{\mu\nu}(n)=\frac{1}{8a^2}\big[Q_{\mu\nu}(n)-Q_{\nu\mu}(n)\big]\ ,
\vspace{0.2cm}
\end{equation}
with
\vspace{0.2cm}
\begin{equation}
\label{Qmunu}
\begin{split}
&Q_{\mu\nu}(n) = U_{\mu}(n)U_{\nu}(n+\hat{\mu})U^{\dagger}_{\mu}(n+\hat{\nu})U^{\dagger}_{\nu}(n) + \\
&+ U_{\nu}(n)U^{\dagger}_{\mu}(n-\hat{\mu}+\hat{\nu})U^{\dagger}_{\nu}(n-\hat{\mu})U_{\mu}(n-\hat{\mu}) + \\
&+ U^{\dagger}_{\mu}(n-\hat{\mu})U^{\dagger}_{\nu}(n-\hat{\mu}-\hat{\nu})U_{\mu}(n-\hat{\mu}-\hat{\nu})U_{\nu}(n-\hat{\nu}) + \\
&+ U^{\dagger}_{\nu}(n-\hat{\nu})U_{\mu}(n-\hat{\nu})U_{\nu}(n+\hat{\mu}-\hat{\nu})U^{\dagger}_{\mu}(n)\ .
\end{split}
\vspace{0.2cm}
\end{equation}

\noindent The perturbative expansion of the $c_{SW}$ coefficient appearing in Eq.(\ref{clover}) is known up to one loop and it can be written as
\vspace{0.1cm}
\begin{equation}
\label{CSW}
c_{SW} = c_{SW}^{(0)} + c_{SW}^{(1)}g_0^2 + \emph{O}(g_0^4)\ ,
\vspace{0.1cm}
\end{equation}
where $c_{SW}^{(0)}=1$~\cite{Sheikholeslami:1985ij} while $c_{SW}^{(1)}$ has been computed in several papers~\cite{Wohlert:1987rf,Luscher:1996vw,Aoki:2003sj,Horsley:2008ap} 
yielding results slightly different but in agreement within errorbars.

Analogously to the case of the gauge action $S_G$, it is worth stressing that, in the present lattice setup featuring SF boundary conditions, an $\emph{O}(a)-$improved 
version of $S_F$ would require not only the addition of $\delta S_V$ as defined in Eq.(\ref{clover}), but also the introduction of boundary counterterms with corresponding 
improvement coefficients to be accurately tuned. Anyway, exactly as it is for the gauge action, such $S_F-$related boundary counterterms will be entirely neglected in this work 
thanks to the fact that the bare PCAC quark mass studied later on is completely fixed by a Ward identity.

Before concluding this section, it is important to observe that the bare quark mass $m_0$ in Eq.(\ref{SF}) will be set to $0$ from now on and that quarks
will be kept massless by subtracting the appropriate mass counterterms order by order in PT (see~\cite{Panagopoulos:2001fn} for their calculation in infinite volume 
to two loops). 

\emph{Methodology -} The improvement coefficient $c_A$ targeted by this study is associated to the isovector axial current $A_{\mu}^b(n)$
\begin{equation}
A_{\mu}^b(n) = \bar{\psi}(n)\gamma_{\mu}\gamma_5\frac{1}{2}\tau^b\psi(n)\ , 
\vspace*{0.2cm}
\end{equation} 
$\tau^b$ being a Pauli matrix acting on flavour indices and $\gamma_5=\gamma_0\gamma_1\gamma_2\gamma_3$ as usual. An $\emph{O}(a)-$improved expression 
is obtained by adding an irrelevant term $\delta A_{\mu}^b(n)$ reading
\vspace{-0.05cm}
\begin{equation}
\label{conA}
\delta A_{\mu}^b(n) = ac_A\frac{1}{2}(\delta_{\mu}^*+\delta_{\mu})P^b(n)\ ,
\vspace{0.2cm}
\end{equation}
where $\delta_{\mu}^*$ and $\delta_{\mu}$ stand for the standard left and right derivative on the lattice while $P^b(n)$ is the isovector axial density
\vspace{0.1cm}
\begin{equation}
P^b(n) = \bar{\psi}(n)\gamma_5\frac{1}{2}\tau^b\psi(n)\ .
\end{equation}
As in Eq.(\ref{CSW}), the improvement coefficient $c_A$ in Eq.(\ref{conA}) can be expanded as
\vspace{0.2cm}
\begin{equation}
\label{CA}
c_A = c_{A}^{(0)} + c_{A}^{(1)}g_0^2 + c_{A}^{(2)}g_0^4 + \ldots\ ,
\vspace{0.2cm}
\end{equation} 
where $c_{A}^{(0)}$ is equal to $0$~\cite{Heatlie:1990kg}, $c_A^{(1)}$ has been determined in~\cite{Luscher:1996vw} while estimating $c_A^{(2)}$ is the goal of this work.

Following~\cite{Luscher:1996sc}, we begin by relating $A_{\mu}^b(n)$ and $P^b(n)$ to the unrenormalized PCAC quark mass $m$ by means of the PCAC relation
\vspace{0.2cm}
\begin{equation}
\label{PCAC}
\Big\langle\frac{1}{2}(\delta_{\mu}^*+\delta_{\mu})A_{\mu}^b(n)\mathcal{O}\Big\rangle = 2m\Big\langle P^b(n)\mathcal{O}\Big\rangle\ ,
\vspace{0.2cm}
\end{equation}
with $\mathcal{O}$ the product of fields located at non-zero distance from site $n$ and from each other. Then, we set $\mathcal{O}$ to be 
\vspace{0.1cm}
\begin{equation}
\label{OP}
\mathcal{O} = a^6\sum_{n',n''}\bar{\zeta}(n')\gamma_5\frac{1}{2}\tau^b\zeta(n'')\ ,
\vspace{0.1cm}
\end{equation}
with the constraint $n'_0=n''_0=0$ and with
\vspace{0.2cm}
\begin{equation}
\left.\zeta(n)\right|_{n_0=0} = \frac{\delta}{\delta\bar{\rho}(\vec{n})}, \ \ \ \ \ \left.\bar{\zeta}(n)\right|_{n_0=0} = -\frac{\delta}{\delta\rho(\vec{n})}\ ,
\vspace{0.2cm}
\end{equation}
being $\rho(\vec{n})$ and $\bar{\rho}(\vec{n})$ the fields introduced in Eqs.(\ref{DirF1})(\ref{DirF2}). 

After defining the correlators $f_A(n)$ and $f_P(n)$ as
\vspace{0.2cm}
\begin{equation}
\begin{split}
\label{ff}
f_A(n) = -a^6\sum_{n',n'',b}\frac{1}{3}\Big\langle A_0^b(n)\bar{\zeta}(n')\gamma_5\frac{1}{2}\tau^b\zeta(n'')\Big\rangle\ , \\
f_P(n) = -a^6\sum_{n',n'',b}\frac{1}{3}\Big\langle P^b(n)\bar{\zeta}(n')\gamma_5\frac{1}{2}\tau^b\zeta(n'')\Big\rangle\ ,
\end{split}
\vspace{0.2cm}
\end{equation}
again with the constraint $n'_0=n''_0=0$, the unimproved bare PCAC quark mass $m$ in Eq.(\ref{PCAC}) is given by
\vspace{0.2cm}
\begin{equation}
\label{mqunimp}
m = \frac{1}{2}\Big[\frac{1}{2}(\delta_{0}^*+\delta_{0})f_A(n)\Big]/f_P(n)\ .
\vspace{0.2cm}
\end{equation}
Noting that $P^b(n)$ is already $\emph{O}(a)-$improved~\cite{Luscher:1996sc} and recalling Eq.(\ref{conA}), the 
$\emph{O}(a)-$improved bare PCAC quark mass $m_{imp}$ is given by
\vspace{0.2cm}
\begin{equation}
\label{mqimp}
m_{imp} = \frac{1}{2}\Big[\frac{1}{2}(\delta_{0}^*+\delta_{0})f_A(n)+ac_A\delta_0^*\delta_0f_P(n)\Big]/f_P(n)\ ,
\vspace{0.2cm}
\end{equation}
provided that the irrelevant term in Eq.(\ref{clover}) is also added to $S_F$ and that both $c_A$ and $c_{SW}$ are correctly set.

The last observation gives us a prescription to determine the improvement coefficients. Before explaining why, it is worth recalling that, in order to study the
continuum limit of a given observable to monitor $\emph{O}(a)$ effects, such an observable must necessarily be a meaningful dimensionless quantity. In this respect,
the most straightforward observable that can be built in the present case is given by the product of the lattice extent $L_S$ times the \emph{renormalized} improved PCAC quark 
mass $m_R$, where 

\vspace{0.2cm}
\begin{equation}
\label{mR}
m_R = \frac{Z_A}{Z_P}m_{imp}\ ,
\vspace{0.2cm}
\end{equation}
$Z_A$ and $Z_P$ being the renormalization constants of the isovector axial current and density respectively. However, it is possible to show explicitly up to 2-loop order that
the multiplicative renormalization of the PCAC quark mass as well as the renormalization of the gauge coupling can eventually be omitted for our purposes, as they only introduce 
$\emph{O}(a^2)$ corrections. This provided of course that the bare quark mass $m_0$ in Eq.(\ref{SF}) is adjusted to its critical value and that $c_A$ is properly set up to 1-loop 
order. In other words, if such a setup holds, $c_A$ can be determined up to the second loop by studying the behaviour of the product of the \emph{unrenormalized} improved PCAC 
quark mass $m_{imp}$ times $L_S$.

Bearing these observations in mind, in PT the product $m_{imp}L_S$ can be expanded as
\vspace{0.1cm}
\begin{equation}
\label{mqimp2}
m_{imp}L_S = \sum_{i=0} m_{imp}^{(i)}L_Sg_0^{2i}\ ,
\vspace{-0.1cm}
\end{equation} 
where coefficients $m_{imp}^{(i)}$ will depend on the coefficients $c_A$ and $c_{SW}$ and on the kinematic parameters $a$, $N_S$, $N_T$, $\theta$ as well as on the 
time coordinate $n_0$ of site $n$ in Eq.(\ref{mqimp}) --- there is no dependence with respect to the spatial coordinates of $n$ because of the translational invariance 
along the corresponding directions. $m_{imp}$ should also depend on the boundary fields $C_k,C_k',\rho,\bar{\rho},\rho',\bar{\rho}'$\footnote{It is worth stressing 
that the dependence of the bare PCAC quark mass on both the kinematical parameters and, in particular, the boundary fields referred to below Eq.(\ref{mqimp2}) is a pure 
lattice artifact. In the continuum, such mass is solely determined by a Ward identity.}: however, fermionic boundary fields will be set to zero after derivatives in 
Eq.(\ref{OP}) are computed while fields $C_k(\vec{n})$ and $C_k'(\vec{n})$ will be fixed to $0$ for every $\vec{n}$ (so that $W_k(\vec{n})=W_k'(\vec{n})=1$). The last 
choice will be motivated later on. 

Since infinite-volume mass counterterms are subtracted up to two loops and since the product $m_{imp}L_S$ does not carry any dimension, both 
$m_{imp}^{(i)}L_S$ --- with $i=1,2$ --- can be expanded in $a$ out of dimensional analysis as
\vspace{0.4cm}
\begin{equation}
\label{mhata}
\begin{split}
m_{imp}^{(i)}L_S &= d_{1,N_S}^{\ \!\!(i)}\frac{a}{L_S} + d_{1,N_T}^{\ \!\!(i)}\frac{a}{L_T} + \\
&+d_{1,\theta}^{\ \!\!(i)}\frac{a\theta}{L_S} + d_{1,n_0}^{\ \!\!(i)}\frac{a}{an_0} + \ldots \ ,\\
&= d_{1,N_S}^{\ \!\!(i)}\frac{1}{N_S} + d_{1,N_T}^{\ \!\!(i)}\frac{1}{N_T-1} + \\
&+ d_{1,\theta}^{\ \!\!(i)}\frac{\theta}{N_S} + d_{1,n_0}^{\ \!\!(i)}\frac{1}{n_0} + \ldots \ ,\\
&
\end{split}
\vspace{0.2cm}
\end{equation}
where dots denote terms of higher order in $a$ and all coefficients $d_{1,\ldots}^{\ \!\!(i)}$ depend on $c_A$ and $c_{SW}$ --- such dependence will be left implicit to ease 
the notation. By setting $N_T=2N_S+1$, $n_0=N_S/2$ (as in~\cite{Luscher:1996vw}) and by keeping $\theta$ fixed, the resulting mathematical setup is such that the terms on the 
r.h.s. of the previous formula can be collected into one as
\vspace{0.1cm}
\begin{equation}
\label{mhat}
m_{imp}^{(i)}L_S = d^{\ \!\!(i)}_{1}\frac{1}{N_S} + O\bigg(\frac{1}{N_S^2}\bigg) .
\vspace{0.1cm}
\end{equation}
\noindent By comparing Eqs.(\ref{mhata}) and (\ref{mhat}), it should be evident that an expansion in powers of $a$ is equivalent to an expansion in powers of $1/N_S$.
Bearing this observation in mind and recalling that we aim at $\emph{O(a)-}$improvement, coefficients $c_{SW}^{(i)}$ and $c_A^{(i)}$ in Eqs.(\ref{CSW}) and (\ref{CA}) 
can be determined up to two loops as follows: with the setup outlined above, the product $L_S$ times the bare PCAC quark mass is first measured for several values of $N_S$, 
then it is fitted vs. $1/N_S$ and $c_{SW}$ and $c_A$ are finally determined by requiring the coefficient $d^{\ \!\!(i)}_{1}$ in Eq.(\ref{mhat}) to be compatible with zero.

In this approach, there is actually one last issue to be solved: in fact, to a given loop $i$ in PT, the coefficient $d^{\ \!\!(i)}_{1}$ depends on all $c_A^{(j)}$ and 
$c_{SW}^{(j)}$ with $j\le i$, so that their effects have to be disentangled. This can be done by choosing the boundary fields $C_k$ and $C_k'$ appropriately. In particular, 
if such fields are both set to $0$ everywhere along the time boundaries as in the present setup, it can be proven~\cite{Luscher:1996sc} that, at the lowest order in PT, the 
dynamical gauge degrees of freedom $U$ are $1$ throughout the whole lattice and, consequently, Eqs.(\ref{Fmunu}) and (\ref{Qmunu}) imply\footnote{This result can be obtained 
with some algebra after the introduction of the formal perturbative expansion in $g_0$ described in the next section.} that the lowest order of $F_{\mu\nu}$ in 
Eq.(\ref{clover}) will be proportional to $g_0$. In turn, this means that, truncating any expansion in $g_0$ at a given loop $i$, only $c_A^{(i)}$ --- as well as all 
coefficients at loops lower than $i$ in Eqs.(\ref{CSW}) and (\ref{CA}) --- will be left into play. This yields to a well-defined procedure to evaluate $c_A^{(2)}$: in fact, 
in the present situation where $c_{SW}^{(i)}$ and $c_A^{(i)}$ have already been determined for $i\le 1$, by setting these tree-level and one-loop coefficients to their known 
values as well as the fields $C_k$ and $C_k'$ to $0$ and by truncating any perturbative expansion at the second loop in $g_0$, $d^{\ \!\!(2)}_{1}$ will only depend on 
$c_A^{(2)}$ and the latter coefficient can thus be fixed by fitting $m_{imp}^{(2)}L_S$ with respect to $1/N_S$. 

Though it is not the goal of this study, let us recall how $c_{SW}^{(2)}$ could be evaluated. The very same setup needed to compute $c_A^{(2)}$ is maintained 
but the fields $C_k$ and $C_k'$ have now to be set as explained in Sect. $6.2$ of~\cite{Luscher:1996sc}: fixing $c_A^{(2)}$ to the value found as outlined in the previous 
paragraphs, $d^{\ \!\!(2)}_{1}$ will now depend solely on $c_{SW}^{(2)}$, so that the correct value of the latter coefficient could be determined, again by fitting 
$m_{imp}^{(2)}L_S$ vs. $1/N_S$. This overall procedure can obviously be iterated to the third loop (and higher), provided that the corresponding mass counterterm is 
subtracted.

Before concluding this section, it is worth recalling that, within the Schr\"odinger Functional formalism, also boundary irrelevant terms in $a$ have in principle to be 
introduced in order to achieve $\emph{O}(a)-$improvement, each one with its own coefficient. However, as stated in~\cite{Luscher:1996vw}, these terms can be eventually 
dropped and remaining improvement coefficients can be determined by solely requiring the unrenormalized PCAC quark mass to be independent of the kinematic parameters, which 
corresponds to the strategy outlined above. 

\emph{NSPT practice -} In this section we describe how configurations are generated by means of Numerical Stochastic Perturbation Theory. NSPT stems from Stochastic 
Quantization (SQ)~\cite{Parisi:1980ys}, a quantization prescription that, in turn, inspired the so-called Langevin algorithm (described in what follows) allowing for the 
computation of expectation values in quantum field theories. It has been used in several domains of research, one of the latest being the search for solutions to the sign 
problem --- see \cite{Aarts:2017vrv,DiRenzo:2015foa} and references therein.

To introduce the basics of SQ in a simple way, we start with a lattice scalar field theory with action $S[\phi]$. In SQ its degrees of freedom $\phi(n)$ are updated by 
numerically integrating a Langevin equation reading 
\vspace{0.12cm}
\begin{equation}
\label{Langevin}
\frac{\partial \phi(n,t)}{\partial t} = -\frac{\partial S[\phi]}{\partial \phi(n,t)} + \eta(n,t)\ ,
\vspace{0.1cm}
\end{equation}
where $t$ is the so-called stochastic time and $\eta(n,t)$ is a Gaussian noise satisfying
\vspace*{0.2cm}
\begin{equation}
\label{scalnoise}
\begin{split}
\langle\eta(n,t)\rangle_{\eta} &= 0\ ,\\
\langle\eta(n,t)\eta(n',t')\rangle_{\eta} &= 2\delta(n-n')\delta(t-t')\ .\\
&
\end{split}
\end{equation}
\noindent The subscript ``$\eta$'' stands for an average over the noise. Given a generic observable $O(\phi)$, it can be shown~\cite{Floratos:1982xj} that the time average
\vspace{0.1cm}
\begin{equation}
\bar{O}(\phi) = \lim_{T\rightarrow+\infty}\frac{1}{T}\int_0^T\!\!dt\ \!O(\phi)\ ,
\vspace{0.1cm}
\end{equation}  
is equal to the path-integral mean value, i.e.,
\vspace{0.2cm}
\begin{equation}
\label{equiv}
\bar{O}(\phi) = \frac{1}{Z}\int D\phi\ \!O(\phi)\ \!e^{-S[\phi]}\ ,
\vspace{0.2cm}
\end{equation}
$Z$ being the partition function. After discretizing the stochastic time $t$ as $t=m\epsilon$ (with integer $m$), Eq.(\ref{Langevin}) can be numerically 
integrated through the prescription\footnote{In what follows, the dependence of any degree of freedom with respect to the discretized stochastic time will
be left implicit, unless needed: in this case, only the integer index ``m'' will be retained and the time step $\epsilon$ will be dropped.}
\vspace{0.2cm}
\begin{equation}
\label{numint1}
\phi(n,m+1) = \phi(n,m) - f(n,m)\ , 
\vspace{0.2cm}
\end{equation}
where the force term $f(n,m)$ in the Euler scheme is given by
\begin{equation}
\label{numint2}
f(n,m) = \epsilon\frac{\partial S[\phi]}{\partial\phi(n,m)} - \sqrt{\epsilon}\ \!\eta(n,m)\ ,
\vspace{0.1cm}
\end{equation}
with $\eta(n,m)=\sqrt{\epsilon}\ \!\eta(n,t=m\epsilon)$. Since the equivalence in Eq.(\ref{equiv}) holds only for continuous $t$, computer simulations with different 
values of $\epsilon$ have to be carried out to extrapolate to $\epsilon\rightarrow 0$. It is worth stressing that an accept/reject step after each update\footnote{At 
present, within SQ there is actually no prescription to implement such an accept/reject step using Runge-Kutta (RK) integrators. However, an exact SQ-inspired algorithm can be 
implemented as a variant of the Generalized HMC algorithm, as long as the approach is non-perturbative. In fact, as stressed in the main text, as soon as PT is introduced, 
in the SQ framework no strategy to implement an accept/reject step is known, irrespective of whether RK integrators or variants of the HMC algorithm are used.} would make 
the algorithm exact and, consequently, no extrapolation in $\epsilon$ would be needed any more since no step-size error would be left into play. Unfortunately, implementing 
an accept/reject step in the NSPT setup --- introduced later on in this section --- is not straightforward because of the perturbative character intrinsic to NSPT itself. 
Therefore, in the NSPT framework, step-size errors can be eliminated only by extrapolating to $\epsilon\rightarrow 0$, i.e., by performing simulations with different values 
of $\epsilon$. However, taking into account the overall scarcity of algorithms allowing for numerical computations within PT, we deem said drawback of NSPT as altogether 
mild and, consequently, consider NSPT a valuable tool to tackle lattice studies in a perturbative framework.  

The SQ setup for the scalar theory has to be modified in order to be applied to $SU(3)$ link variables. In this respect, Eq.(\ref{Langevin}) is modified as
\vspace{0.1cm}
\begin{equation}
\label{LangU}
\frac{\partial U_{\mu}(n,t)}{\partial t} = -i\sum_{a=1}^8T^a[\nabla_{\!n,\mu}^a S_G[U] - \eta_{\mu}^a(n,t)] U_{\mu}(n,t)\ , 
\end{equation}
where matrices $T^a$ are the generators of the $SU(3)$ algebra (with normalization $tr(T^aT^b)=\frac{1}{2}\delta^{ab}$) and $\nabla_{n,\mu}^a$ is the Lie derivative 
--- with respect to the algebra fields associated to variable $U_{\mu}(n)$ --- defined as~\cite{Drummond:1982sk}
\vspace{0.2cm}
\begin{equation}
f[e^{i\sum_a\!\omega^aT^a}U] = f[U] + \sum_{a}\omega^a\nabla^a\!f[U] + \emph{O}(\omega^2)\ ,
\vspace{0.2cm}
\end{equation}
$f[U]$ being a scalar function of the group variable $U$ and $\omega^a$'s small parameters. The noise $\eta_{\mu}^a(n,t)$ appearing in Eq.(\ref{LangU}) satisfies the 
conditions
\vspace*{0.1cm}
\begin{equation}
\label{groupnoise}
\begin{split}
\langle\eta_{\mu}^a(n,t)\rangle_{\eta} &= 0\ ,\nonumber\\
\langle\eta_{\mu}^a(n,t)\eta_{\nu}^b(n',t')\rangle_{\eta} &= 2\delta(n-n')\delta(t-t')\delta_{\mu\nu}\delta^{ab}\ ,\\
\end{split}
\end{equation}
i.e., the straightforward extension of Eq.(\ref{scalnoise}) incorporating the degrees of freedom associated to space-time directions and group components.

The group counterpart of Eq.(\ref{numint1}) reads
\vspace{0.2cm}
\begin{equation}
\label{update}
U_{\mu}(n,m+1) = e^{-i\sum_a\!T^a\!f^a_{\mu}(n,m)}U_{\mu}(n,m)\ ,
\vspace{0.2cm}
\end{equation}
with
\vspace{0.2cm}
\begin{equation}
\label{update2}
f^a_{\mu}(n,m) = \epsilon\nabla_{\!n,\mu}^aS_G[U] - \sqrt{\epsilon}\ \!\eta^{a}_{\mu}(n,m)\ .
\vspace{0.2cm}
\end{equation}

In this framework, PT up to $N_{L}$ loops can be introduced by a formal expansion of each gauge field as
\vspace*{0.4cm}
\begin{equation}
\begin{split}
\label{PTAU}
A_{\mu}(n) &= A^{(1)}_{\mu}(n)\beta^{-\frac{1}{2}} + A^{(2)}_{\mu}(n)\beta^{-1} + \\ 
&+ \ldots + A^{(2N_L)}_{\mu}(n)\beta^{-N_L}\ ,\\
U_{\mu}(n) &= 1 + U^{(1)}_{\mu}(n)\beta^{-\frac{1}{2}} + U^{(2)}_{\mu}(n)\beta^{-1} + \\
&+ \ldots + U^{(2N_L)}_{\mu}(n)\beta^{-N_L}\ ,\\
&
\end{split}
\end{equation}
with $\beta = 6/g_0^2$. A few remarks are in order concerning Eqs.(\ref{PTAU}). First, the leading order of $U_{\mu}(n)$ is $1$ in the light of the previous choice
of the fields $C_k$ and $C_k'$ (see the comments at the end of the previous section). Second, while the fields $A^{(i)}_{\mu}(n)$ are elements of the Lie 
Algebra of $SU(3)$, the fields $U_{\mu}^{(i)}(n)$ --- \emph{taken one by one} --- do not belong to the $SU(3)$ group but $U_{\mu}(n)$ on the l.h.s. of the second of 
Eqs.(\ref{PTAU}) does, excluding terms of order $\emph{O}(\beta^{-N_L-1/2})$ and higher in PT. Finally, a Taylor expansion of the exponential of $A_{\mu}(n)$ allows to 
obtain the group variable $U_{\mu}(n)$, while a similar expansion of the logarithm of $U_{\mu}(n)$ results in $A_{\mu}(n)$.

Plugging the second of Eqs.(\ref{PTAU}) into a discretized version of Eq.(\ref{LangU}) results in a hierarchical system of differential equations where the evolution
of a given order $U^{(i)}_{\mu}(n)$ only depends on lower orders, thus allowing for a consistent truncation at the needed loop $N_L$. In this setup, the noise 
$\eta^{a}_{\mu}(n)$ enters at order $g_0$ after a further rescaling of $\epsilon$ with $\beta$, as explained in \cite{DiRenzo:1995qc}. 

This is the core of the NSPT algorithm which has been applied to lattice QCD in order to study --- among others --- the free energy density at finite 
temperature~\cite{DiRenzo:2004ws,DiRenzo:2006nh,DiRenzo:2008en}, renormalization constants~\cite{DiRenzo:2006qtj,DiRenzo:2009ni,DiRenzo:2010cs,Brambilla:2013sua,
Brambilla:2014bja} and renormalons~\cite{DiRenzo:1994sy,DiRenzo:1995qc,DiRenzo:2000ua,Bali:2013pla,Bali:2014fea,Bali:2014sja,DelDebbio:2018ftu}.

In order to prevent fluctuations associated to the random-walk behaviour of gauge modes~\cite{DiRenzo:2004hhl}, the updating step in Eq.(\ref{update}) has to be alternated 
with the so-called stochastic gauge fixing~\cite{Zwanziger:1981kg}. In other words, before moving from configuration $U_{\mu}(n,m)$ to 
$U_{\mu}(n,m+1)$ in stochastic time, each field $U_{\mu}(n,m)$ has to undergo a gauge transformation like
\vspace*{0.2cm}
\begin{equation}
U_{\mu}(n,m)\rightarrow G(n,m)U_{\mu}(n,m)G^{\dagger}(n+\hat{\mu},m)\ ,
\vspace*{0.1cm}
\end{equation} 
where matrices $G$ belong to the $SU(3)$ group as well. With periodic boundary conditions along all directions, a common choice for $G(n)$ is $G(n)=exp[\Omega(n)]$ 
where $\Omega(n)$ reads~\cite{Rossi:1987hv}
\vspace*{0.1cm}
\begin{equation}
\Omega(n) = -i\alpha\sum_{\mu}\partial^*_{\mu}A_{\mu}(n)\ \ \ \ \ (0<\alpha<1)\ ,
\end{equation}
which results in fluctuations around the Landau gauge. With Dirichlet boundaries, this definition of the entries of matrix $\Omega(n)$ is modified~\cite{Brambilla:2013sba} 
for sites with 
$n_0=0$ or $n_0=N_T-1$ as follows:
\vspace*{0.4cm}
\begin{equation}
\Omega(n)_{ab} = \left\{
\begin{array}{ll}
\frac{1}{N_S^3}\sum_{n'} [A_0(n')]_{ab} & \text{if } a=b \text{ and } n_0=0\ , \nonumber\\
0 & \text{otherwise}\ ,
\end{array}\right.
\vspace*{0.3cm}
\end{equation}
where the constraint $n'_0=0$ holds in the sum over $n'$.

To carry out the computation of correlators $f_A(n)$ and $f_P(n)$ entering in the bare PCAC quark mass, it is necessary to contract the fermionic fields in 
Eqs.(\ref{ff}) and, therefore, to invert the Wilson-Dirac operator $D$ in Eq.(\ref{SFunimp}). This is done following~\cite{DiRenzo:2004hhl}: given a generic 
operator $M$ and its perturbative expansion $M = \sum_{k=0} g_0^kM^{(k)}$, its inverse $M^{-1}$ can be expanded perturbatively as
\vspace*{0.2cm}
\begin{equation}
M^{-1} = M^{(0)^{-1}} + \sum_{k>0}g_0^kM^{-1^{(k)}}\ ,
\vspace*{0.2cm}
\end{equation}
where the tree-level term is the inverse of the zero-order term of $M$ while, recursively, 
\vspace*{0.2cm}
\begin{equation}
\begin{split}
M^{-1^{(1)}} &= -M^{(0)^{-1}}M^{(1)}M^{(0)^{-1}}\ ,\nonumber\\
M^{-1^{(2)}} &= -M^{(0)^{-1}}M^{(2)}M^{(0)^{-1}} + \nonumber\\
&-M^{(0)^{-1}}M^{(1)}M^{{-1}^{(1)}}\ ,\nonumber\\
\ldots & \nonumber\\
M^{-1^{(n)}} &= -M^{(0)^{-1}}\sum_{j=0}^{n-1}M^{(n-j)}M^{{-1}^{(j)}}\ .
\end{split}
\end{equation}
The expression for $D^{(0)^{-1}}$ can be found in Sect. 3.1 of~\cite{Luscher:1996vw}.

Finally, it is worth mentioning that, in the setup discussed so far, there are no zero modes to take care of: this is due the Dirichlet boundary conditions for the
fermions and to the non-vanishing value of the angle $\theta$ associated to the covariant derivatives in Eq.(\ref{covdev}) --- the latter condition actually modifies 
(typically increases) the spectral gap of the massless Dirac operator with SF boundaries.

\begin{figure*}[ht!]
\includegraphics[width=0.49\textwidth,height=5.5cm]{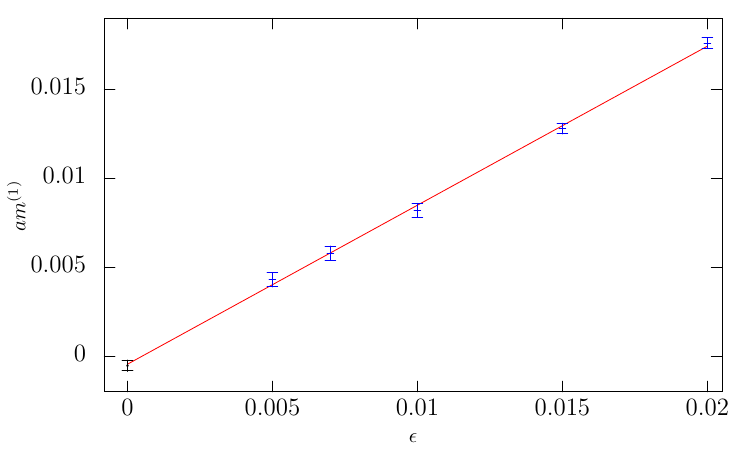}
\includegraphics[width=0.49\textwidth,height=5.5cm]{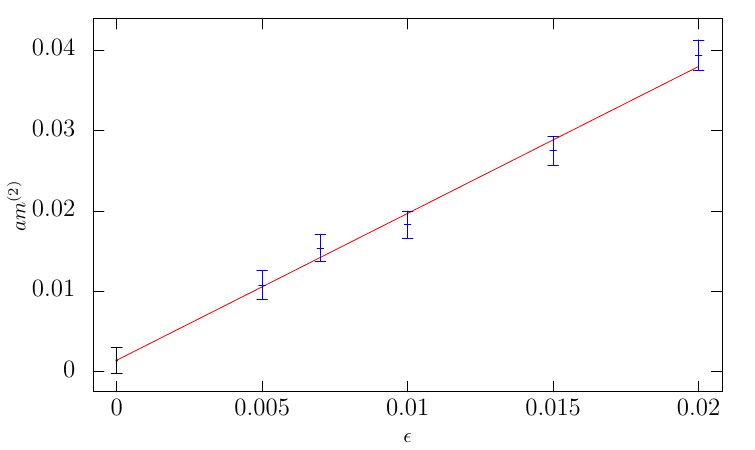}
\caption{(Color online) An example of the extrapolation to $\epsilon\rightarrow 0$ of the unimproved bare PCAC quark mass in lattice units at the first (left) and second loop 
(right) in powers of $\beta^{-1/2}$ with $N_S=24$. In both panels the black, leftmost point represents the extrapolated result, the points in blue the results at finite $\epsilon$ 
obtained with the NSPT simulations while the red, continuous line corresponds to the fit function.}
\label{epsext}
\end{figure*}

\begin{table}[t]
\begin{tabular}{c c r c r c r c r c r}
$N_S$   & &
\multicolumn{9}{c}{$\epsilon$} \\
\hline
   & & \ 0.005 & & \ 0.007 & & \ 0.010 & & \ 0.015 & & \ 0.020 \\
\hline
11 & & 21.2 & & 19.8 & & 19.7 & & 19.7 & & 19.7 \\   
12 & & 20.4 & & 18.0 & & 25.2 & & 19.7 & & 18.9 \\   
13 & & 19.8 & & 21.8 & & 19.8 & & 24.8 & & 19.3 \\   
14 & & 19.0 & & 29.9 & & 27.9 & & 28.1 & & 28.2 \\   
15 & & 18.5 & & 18.0 & & 18.0 & & 19.5 & & 18.0 \\   
16 & & 17.9 & & 18.1 & & 18.0 & & 17.9 & & 18.8 \\   
17 & & 18.8 & & 18.6 & & 18.9 & & 18.3 & & 17.9 \\   
18 & & 17.3 & & 15.8 & & 17.6 & & 13.3 & & 17.0 \\   
20 & & 13.5 & & 13.1 & & 15.4 & & 13.3 & & 12.8 \\   
24 & & 12.6 & & 13.3 & & 12.8 & & 13.3 & & 13.4 \\   
32 & & 3.3 & & 3.3 & & 3.0 & & 2.8 & & 2.6 \\   
\end{tabular}
\caption{Number of measurements (in tens of thousands) of $f_A$ and $f_P$ for the different combinations of simulation parameters ($N_S$,$\ \!\epsilon$).}
\label{tabNeps}
\end{table}

\begin{table}[t]
\begin{tabular}{c c r c r c c c r}
$N_S$ & & $am^{(1)}$\ \ \ & & $am^{(2)}$\ \ \ \ & & $a\delta m^{(0)}$\! & & $a\delta m^{(1)}$ \ \ \!\\
\hline
11 & & 0.0031(2) & & 0.0048(04) & & 0.0711 & & -0.0460(65) \\
12 & & 0.0029(3) & & 0.0021(16) & & 0.0598 & & -0.0494(64) \\
13 & & 0.0025(5) & & 0.0018(28) & & 0.0510 & & -0.0408(51) \\
14 & & 0.0018(3) & & 0.0032(18) & & 0.0440 & & -0.0307(67) \\
15 & & 0.0013(2) & & 0.0062(21) & & 0.0383 & & -0.0192(55) \\
16 & & 0.0011(5) & & 0.0006(10) & & 0.0337 & & -0.0134(25) \\
17 & & 0.0017(6) & & 0.0043(32) & & 0.0298 & & -0.0148(60) \\
18 & & 0.0012(3) & & 0.0003(14) & & 0.0266 & & -0.0083(17) \\
20 & & 0.0002(5) & & 0.0039(17) & & 0.0216 & & -0.0013(37) \\
24 & & -0.0005(3) & & 0.0014(16) & & 0.0150 & & -0.0006(31) \\
32 & & 0.0017(7) & & -0.0003(38) & & 0.0084 & & 0.0060(46) \\
\end{tabular}
\caption{$\epsilon\rightarrow0$ results for the one- and two-loop contribution $m^{(1)}$ and $m^{(2)}$ to the unimproved PCAC quark mass and for the improvement mass term
at tree level ($\delta m^{(0)}$) and first-loop order ($\delta m^{(1)}$) for the values of $N_S$ used in this study: all these quantities are here expressed in lattice units. 
The values in brackets correspond to the statistical error affecting all observables but those at tree level as explained in footnote 8. The perturbative expansion all these 
observables are associated to is \emph{in powers of $\beta^{-1/2}$}.}
\label{tabmphysunits}
\end{table}

\emph{Data analysis and results -} Simulations were performed with the following values of the parameters: $\theta=1.2$, $\epsilon\in\{0.005,0.007,0.010,0.015,0.020\}$ 
and $N_S\in\{11,12,13,14,15,16,17,18,20,24,32\}$. Note that, since the Euler scheme is employed in the integration of the Langevin equation, only three values of 
$\epsilon$ would be actually needed to extrapolate to $\epsilon\rightarrow 0$. However, we used $5$ time steps in order to increase the precision of the extrapolation. 
In our analysis we did not make use of the data obtained from the simulations featuring $N_S$ lower than $11$ since such data turned out to be rather noisy --- 
this phenomenon is actually puzzling and, unfortunately, we have to admit that we could not find any convincing explanation for it.\\
\indent Tab.(\ref{tabNeps}) details the number of measurements of observables $f_A$ and $f_P$ in Eqs.(\ref{ff}) --- and, hence, of the unrenormalized PCAC quark mass --- 
for the different combinations of simulation parameters $(N_s,\epsilon)$. For each setup two subsequent measurements were separated by 100 Langevin updates of the lattice 
configuration in order to reduce the autocorrelation\footnote{In order to give the reader an idea of the simulation cost, we quote the autocorrelation time ACT of the most 
expensive observable that has been measured, i.e. the two-loop contribution to the bare PCAC quark mass at $N_S=32$ and $\epsilon=0.005$: for said observable in such a setup, 
the ACT reads approximately 200 Langevin updates.}. It is worth stressing that these 100 updating steps usually took approximately half of the time needed to perform a single 
measurement of the bare PCAC quark mass.\\
\indent Fig.(\ref{epsext}) shows a plot of the unimproved bare PCAC quark mass in lattice units --- see Eq.(\ref{mqunimp}) for its definition in physical units --- at the first 
($am^{(1)}$) and second loop ($am^{(2)}$) in powers of $\beta^{-1/2}$ vs. $\epsilon$ together with the extrapolated result. Statistical errors on the data in blue in 
Fig.(\ref{epsext}) have been obtained through a jackknife procedure.\\
\indent Before computing $c_A^{(2)}$, we check to which extent our approach is reliable: it is understood that all bare PCAC quark masses referred to in the rest of this 
section are obtained from an extrapolation to $\epsilon\rightarrow0$ as that shown in Fig.(\ref{epsext})\footnote{The only exception to this assumption is given by 
tree-level quantities since they do not depend on the Langevin time $\epsilon$. This is due to the r.h.s. of Eq.(\ref{update2}), whose leading order is $g_0$: in fact, in 
the present setup, the action derivative is zero at tree level while, as already mentioned before, the Gaussian noise enters at order $g_0$ in PT. Consequently, the tree 
level of any variable $U_{\mu}(n)$ does not evolve with respect to the stochastic time and bears no statistical error.}.\\
\indent As a first test, we compared the analytical result for the tree-level bare PCAC quark mass in lattice units~\cite{Luscher:1996vw} with our estimate. The two values 
agree apart from round-off errors that are usually lower than $0.03\%$ --- the smallest tree-level bare PCAC quark mass we measure is equal to $2.97\cdot10^{-6}$ for $N_S=32$,
 which is also the extent where round-off errors turn out to be the largest.\\
\indent As a second check, we now work out the one-loop coefficient $c_A^{(1)}$ and compare it to the value determined in~\cite{Luscher:1996vw}. To better explain
the fit strategy, we rewrite the product of the lattice extent $L_S$ times the improved bare PCAC quark mass $m_{imp}$ in Eq.(\ref{mqimp}) as
\vspace*{0.3cm}
\begin{equation}
m_{imp}L_S = (m + c_A\delta m)L_S\ ,
\vspace*{0.3cm}
\end{equation}
where $m$ is defined in Eq.(\ref{mqunimp}) and $a$ has been set to $1$ --- let us recall once more that any dependence in the lattice spacing is purely formal. 
Plugging Eq.(\ref{CA}) into the previous formula, at one-loop level we have
\vspace*{0.3cm}
\begin{equation}
\label{m1}
m_{imp}^{(1)}L_S = (m^{(1)} + c_A^{(1)}\delta m^{(0)})L_S\ .
\vspace*{0.3cm}
\end{equation}
$\delta m^{(0)}$ corresponds to the tree-level contribution to the term (including the denominator) multiplied times $c_A$ on the r.h.s. of Eq.(\ref{mqimp}): being a 
quantity at tree level, it bears no error.

In order to determine $c_A^{(1)}$, we proceed as follows: tentative values $\tilde{c}_A^{\ \!\!(1)}$'s are assigned to $c_A^{(1)}$, $m_{imp}^{(1)}L_S$ is fitted vs. 
$f(1/N_S)=d\frac{1}{N_S}$ and any $\tilde{c}_A^{\ \!\!(1)}$ is eventually retained as valid if the coefficient $d$ is compatible with $0$ within errorbars. In this way, we 
obtain a range $[\tilde{c}_{A,min}^{\ \!\!(1)},\tilde{c}_{A,max}^{\ \!\!(1)}]$ of valid values for $c_A^{(1)}$ and we quote as mean value $\overline{c}_A^{(1)}$ and 
statistical error $\sigma_{c_A^{(1)}}$ on this one-loop improvement coefficient the expressions
\vspace*{0.4cm}
\begin{equation}
\overline{c}_{A}^{(1)} = \frac{\tilde{c}_{A,min}^{\ \!\!(1)}+\tilde{c}_{A,max}^{\ \!\!(1)}}{2}\ , \ \ \ \sigma_{c_{A}^{(1)}} = \frac{\tilde{c}_{A,max}^{\ \!\!(1)}-\tilde{c}_{A,min}^{\ \!\!(1)}}{2}\ .
\vspace*{0.4cm}
\end{equation}

As for the systematic error, its main source is given by the truncation of function $f(1/N_S)$ at the first order in $1/N_S$. Therefore, the best way to assess the 
systematics would be to repeat the procedure above with a function of higher order in $1/N_S$ and to compare the outcome with that obtained with a linear function. In fact, 
as pointed out in~\cite{Luscher:1996vw}, for relatively large values of $\theta$ as that used in this study, corrections going like $1/N_S^2$ can be comparable to the 
leading contribution $1/N_S$. Unfortunately, our data are not precise enough to support higher-order terms in $f(1/N_S)$: any attempt of fit in this sense winds up with 
an extremely poor determination of $c_A^{(1)}$. In order to assess the systematic error in a somehow coarser way, we carried out the fit of $m_{imp}^{(1)}L_S$ vs. 
$f(1/N_S) = d\frac{1}{N_S}$ as before but within a range of $N_S$ limited to the three smallest extents (i.e., $N_S\in[11,13]$). Since the latter is the regime where 
higher-order corrections in $N_S$ should have more impact, the mean value of $c_A^{(1)}$ obtained in this way should feature the largest deviation with respect to the 
mean value of $c_A^{(1)}$ computed with the fit employing all available sizes. Such a deviation could then be considered as a rough estimate of the systematic error.

\begin{figure*}[ht!]
\includegraphics[width=0.49\textwidth,height=6.0cm]{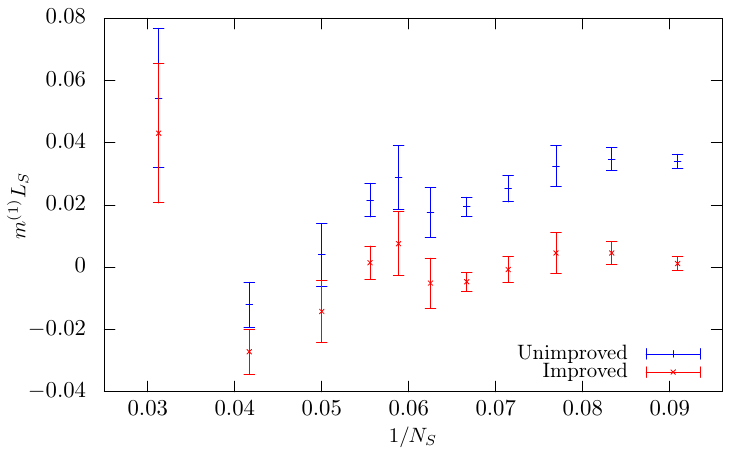}
\includegraphics[width=0.49\textwidth,height=6.0cm]{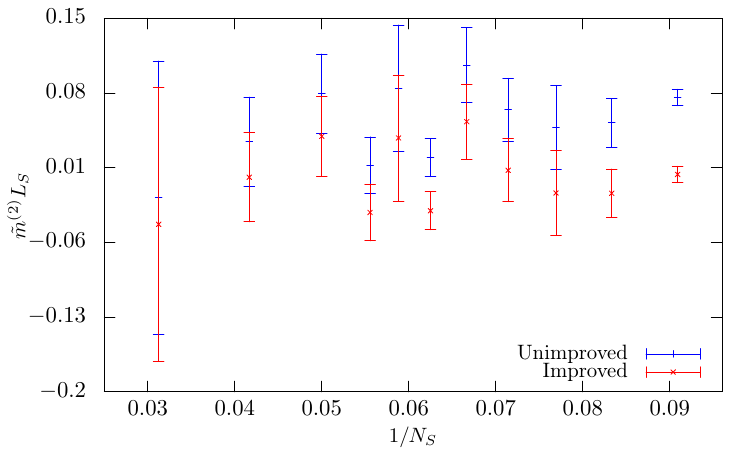}
\caption{(Color online) Plots of $m^{(1)}L_S$ (left, in blue) and $\tilde{m}^{(2)}L_S$ (right, in blue) -- as referred to on the r.h.s. of Eqs.(\ref{m1}) and (\ref{m2}) 
respectively --- vs. $1/N_S$. The points in red denoted with stars correspond to their improved counterparts --- referred to on the l.h.s. of the same equations --- where 
$c_A^{(1)}$ on the left and $c_A^{(2)}$ on the right have been set to the mean values reported in Eqs.(\ref{ca1}) and (\ref{ca2}) respectively (with a conversion from powers 
of $g_0$ to powers of $\beta^{-1/2}$ properly taken into account). Observables in both panels refer to a perturbative expansion in $\beta^{-1/2}$ and can be computed starting 
from the values in Tab.(\ref{tabmphysunits}).}
\label{mass}
\end{figure*}

Following this strategy and converting our expansion in $\beta^{-1/2}$ into a series in $g_0$, the result we obtain for $c_A^{(1)}$ reads
\vspace*{0.2cm}
\begin{equation}
\label{ca1}
c_A^{(1)} = -0.00701(53)(50)\ ,
\vspace*{0.2cm}
\end{equation}
where the first and second error are the statistical and systematic uncertainty respectively. Within errorbars, this values is in reasonable agreement with 
$c_A^{(1)} = -0.00756(1)$ quoted in~\cite{Luscher:1996vw}, though the precision of the latter estimate is much higher than that obtained with NSPT. 

If we now move to the evaluation of $c_A^{(2)}$, we can first write the counterpart of Eq.(\ref{m1}), i.e.,
\vspace*{0.5cm}
\begin{equation}
\begin{split}
\label{m2}
m_{imp}^{(2)}L_S &= (m^{(2)} + c_A^{(1)}\delta m^{(1)} + c_A^{(2)}\delta m^{(0)})L_S = \  \\
&=(\tilde{m}^{(2)} + c_A^{(2)}\delta m^{(0)})L_S\ ,\\
&
\end{split}
\end{equation}
where $c_A^{(1)}$ has been set to its known value $c_A^{(1)} = -0.00756$. Starting from the last expression, the same procedure adopted at one-loop level can be applied to 
the fit of $m_{imp}^{(2)}L_S$ with respect to $1/N_S$, only $m^{(1)}$ in Eq.(\ref{m1}) has to be replaced with $\tilde{m}^{(2)} = m^{(2)} + c_A^{(1)}\delta m^{(1)}$. Similarly to 
what remarked before for $\delta m^{(0)}$, $\delta m^{(1)}$ corresponds to the one-loop contribution to the term (including the denominator) multiplied times $c_A$ on the r.h.s. 
of Eq.(\ref{mqimp}).

After converting again the expansion in $\beta^{-1/2}$ into a series in $g_0$, the result we get for $c_A^{(2)}$ is
\vspace*{0.2cm}
\begin{equation}
\label{ca2}
c_A^{(2)} = -0.00256(22)(6)\ .
\vspace*{0.2cm}
\end{equation}
While at one loop the systematic error is essentially equal to the statistical one, at two-loop level systematic effects seem to be apparently less important. This is most 
likely a consequence of a worse signal-to-noise ratio at the second loop, as shown in Fig.(\ref{mass}) where $m^{(1)}L_S$ and $\tilde{m}^{(2)}L_S$ are plotted -- together with 
their $\emph{O}(a)-$improved counterparts --- on the left and right panel respectively. At one loop, statistical errors are smaller and, consequently, the trend of the data 
is altogether better defined, thus allowing for the rough assessment of subleading corrections in $1/N_S$ outlined above. On the contrary, at two-loop level the trend of 
the data is harder to be determined due to the larger errorbars: this eventually leads to a more difficult estimate of the impact of subleading contributions in $1/N_S$ 
and, in practice, to an apparently small systematics.\\
\indent In order to be somehow more conservative, we assume the systematic error at two-loop level to be roughly comparable to the statistical uncertainty, as it is the case
for the first-loop result in Eq.(\ref{ca1}). In this spirit, the estimate for $c_A^{(2)}$ would read
\vspace*{0.2cm}
\begin{equation}
\label{ca2_cons}
c_A^{(2)} = -0.00256(22)(20)\ .
\vspace*{0.2cm}
\end{equation}

For completeness, in Tab.(\ref{tabmphysunits}) we provide --- for the different values of $N_S$ --- the $\epsilon\rightarrow 0$ results for $m^{(1)}$, $m^{(2)}$, $\delta m^{(0)}$ 
and $\delta m^{(1)}$ (in lattice units), i.e., all the mass contributions entering the computation of $c_A^{(1)}$ and $c_A^{(2)}$ explained in this section.

\emph{Conclusions -} By studying the cutoff dependence of the unrenormalized PCAC quark mass, we carried out an exploratory, perturbative determination of the 
improvement coefficient $c_A$ to two loops by using NSPT in the quenched approximation. The crosschecks at zero- and one-loop in PT reproduce --- with fair precision --- 
results previously obtained in other independent studies, thus indicating that NSPT can in principle be used in this kind of computations, even at orders where
non-numerical approaches are usually too cumbersome from an algebraic point of view. Let us recall that the loop $N_L$ at which the Taylor series in Eqs.(\ref{PTAU}) are
truncated is in fact just a parameter and increasing it does not entail any extra work from the point of view of computer coding. 

\indent However, though viable in theory, at a practical level NSPT seems to be very demanding in terms of computer resources when it comes to producing an accurate 
estimate of the improvement coefficients, even in a relatively simple setup as that examined in this study, featuring the quenched approximation and the Wilson action at 
two loops. Indeed, in spite of having employed some millions of CPU hours, data are still rather noisy, as shown in Fig.(\ref{mass}): though not optimal at the first loop 
already (and, in fact, our estimate for $c_A^{(1)}$ is much less precise than that contained in \cite{Luscher:1996vw}), the deterioration of the signal at higher order in 
PT is manifest and gets particularly reflected in a difficult assessment of the systematic uncertainty. It might be that observables other than that examined in this study 
--- i.e., the unrenormalized PCAC quark mass --- display an intrinsically better signal-to-noise ratio but, at present, we would have no suggestion to make with respect to 
this issue.\\ 
\indent Our understanding is that reducing errorbars to a level at which NSPT data can be accurately fitted (and improvement coefficients precisely determined) would most 
likely require considerable computer resources irrespective of the quantity being monitored, especially if higher order in PT are envisaged and/or more interesting --- 
but also more demanding --- setups are considered, as that of unquenched simulations.\\
\indent More or less recently, some works have actually been published \cite{Luscher:2014mka,DallaBrida:2017pex,DallaBrida:2017tru} proposing some technical changes to the 
standard NSPT approach followed in this study. Such changes yield new setups --- called Instantaneous Stochastic Perturbation Theory (ISPT), Hybrid Stochastic Perturbation 
Theory (HSPT) and Kramers Stochastic Perturbation Theory (KSPT) --- that seem to be highly beneficial in remeding the shortcomings of the standard version of NSPT we 
employed: this looks particularly true for the formulation described in \cite{DallaBrida:2017tru}. It would be extremely interesting to carry out a computation of 
$c_A^{(2)}$ with the latter setup in order to (hopefully) get a more precise estimate of this improvement coefficient at much lower computer costs\footnote{We regret we 
did not experiment with any of these new formulations ourselves. Unfortunately, the first of said papers --- i.e., \cite{Luscher:2014mka} --- was published when the 
present project was almost coming to its end (in fact, personal reasons indipendent of our will considerably delayed the publication of our results). Such a bad timing 
prevented us from taking advantage of any of the techniques described in \cite{Luscher:2014mka,DallaBrida:2017pex,DallaBrida:2017tru}.}.

\indent Nevertheless, we believe that the original result of this project, i.e., the evaluation of the coefficient $c_A^{(2)}$ for $N_f=0$, can serve as a benchmark to some 
extent: indeed, we are confident that at least its negative sign and its order of magnitude (that is, $10^{-3}$) are reliable. 

\emph{Acknowledgements -} We would like to thank G.S. Bali for useful discussions at different stages of this project. Simulations were carried out on the Athene cluster of
the University of Regensburg and on the cluster of the Leibniz-Rechenzentrum in Munich.

\indent This work  is dedicated to my wife, Milena, and to our sons, Edoardo and Martino.


\bibliography{refs}{}
\bibliographystyle{apsrev4-1} 

\end{document}